\documentclass[conference]{IEEEtran}

\usepackage{authblk}
\usepackage{amsmath}
\usepackage{amssymb}
\usepackage{amsthm}
\usepackage{cite}
\usepackage{float}
\usepackage{multirow}
\usepackage{graphicx}
\usepackage{enumerate}
\usepackage{hyperref}
\usepackage{algorithm,algorithmic}
\usepackage{xpatch}
\usepackage{verbatim}
\usepackage{bbm}

\makeatletter
\xpatchcmd\algorithmic
  {\newcommand{\IF}}
  {%
    \newcommand\SCOPE{\begin{ALC@g}}%
    \newcommand\ENDSCOPE{\end{ALC@g}}%
    \newcommand{\IF}%
  }
  {}{\fail}
\makeatother

\hypersetup{
    colorlinks=true,
    linkcolor=blue,
    filecolor=magenta,      
    urlcolor=cyan,
}
\usepackage{tikz}
\usepackage{pgfplots}
\usepgfplotslibrary{colorbrewer}
\pgfplotsset{compat=1.8}
\definecolor{clr1}{RGB}{27,158,119}
\definecolor{clr2}{RGB}{217,95,2}
\definecolor{clr3}{RGB}{117,112,179}
\definecolor{clr4}{RGB}{231,41,138}
\definecolor{clr5}{RGB}{102,166,30}
\definecolor{clr6}{RGB}{230,171,2}
\pgfplotsset{
    cycle list={clr1,clr2,clr3,clr4,clr5,clr6},
}

\begin{document}

\title{Reinforcement learning for Admission Control in 5G Wireless Networks\\

\thanks{
}}

\author[1]{Youri Raaijmakers}
\author[2]{Silvio Mandelli}
\author[2]{Mark Doll}
\affil[1]{Department of Mathematics and Computer Science, Eindhoven University of Technology, The Netherlands}
\affil[2]{Nokia Bell Labs, Germany}

\maketitle

\begin{abstract}
The key challenge in admission control in wireless networks is to strike an optimal trade-off between the blocking probability for new requests while minimizing the dropping probability of ongoing requests. We consider two approaches for solving the admission control problem: i) the typically adopted threshold policy and ii) our proposed policy relying on reinforcement learning with neural networks. Extensive simulation experiments are conducted to analyze the performance of both policies. The results show that the reinforcement learning policy outperforms the threshold-based policies in the scenario with heterogeneous time-varying arrival rates and multiple user equipment types, proving its applicability in realistic wireless network scenarios.
\end{abstract}

\section{Introduction}
\label{sec: introduction}

With the deployment of first releases of 5G 3GPP wireless networks standards, the research community is already defining the scenarios and directions for Beyond 5G and 6G future systems~\cite{6GWhitePaper,harish20206g}.
Interest in Artificial Intelligence techniques has surged, leading to the revolution of many staple techniques commonly used in communications~\cite{hoydis2020special}. At the same time, some of the old challenges remain hot, like enforcing network flexibility with a variety of possible Quality of Service (QoS) demands by its users~\cite{ABKKM-EAMSWA,maeder2016scalable,MABK-SNSC}.

Heterogeneous QoS profiles and increasing traffic demands require the implementation of a proper prioritization framework in base stations. In wireless networks prevention of cell congestion cannot be demanded from real time operations where simpler algorithms, e.g., radio resource scheduling, are running. The typical first layer of security is guaranteed by a proper Admission Control (AC) mechanism.
It can be described as an agent that decides whether or not to accept a new user equipment (UE) connection request, based on the current cell load and the QoS profile of ongoing traffic and of the new request, defacto aiming at preventing congestion and serving as much traffic as possible.

The most relevant performance metrics for AC are the blocking probability, i.e., the probability that a new UE connection request is blocked and the dropping probability, i.e, the probability that an existing UE connection is terminated due to insufficient available resources at the base station. 

The AC problem in cellular systems already received a lot of attention and is well studied~\cite{DM-CALS,RNT-OCAC,MHT-CACCDMA}.
The AC problem for 5G wireless networks is attracting renewed interest because of the new complexities within these networks, see~\cite{AACA-CACUDN, HNLF-ACNS}. Most notable is the state of the art solution using reinforcement learning (RL). Unlike other model-based algorithms, RL does not require specific state transition models which is a very important feature when considering large wireless networks supporting various types of UEs. Especially, an agent based on Q-learning, see~\cite{SBP-DQLAC, TB-ACAC}, and on the use of neural networks, see~\cite{BKO-TASAC, MG-CAC}, is studied. In~\cite{MG-CAC} it is assumed that the channel rate of the UE is constant over time and~\cite{BKO-TASAC} considers the scenario with stochastic resource requirements network slices that are fixed over time. 

As alluded to above, the goal of the AC agent is to strike an optimal trade-off between the blocking and dropping probability, all in the presence of varying channel rates of the UE. 
The novelty of our approach compared to the closest prior art~\cite{BKO-TASAC, MG-CAC} is the consideration of UE mobility that impacts the large scale components of the wireless channel, according to~\cite{TR38.901}.
So it may be that, at the time of a new UE connection request, the base station has enough available resources to serve it. However, due to varying channel rates given by different (i) UE position and (ii) serving cell, the required resources to guarantee the same throughput QoS may change. Therefore, a UE connection may be dropped since the base station is not capable of providing enough resources to all the UEs.
We consider two approaches for solving the AC problem: i) a threshold policy and ii) a reinforcement learning policy. Extensive simulation experiments are conducted to analyze the performance of both policies. The results show that the reinforcement learning policy outperforms the threshold-based policies, being able to generalize its operations in the scenario with heterogeneous time-varying arrival rates and multiple UE types.

The remainder of the paper is organized as follows. In Section~\ref{sec: methods} we explain the AC policies. In Section~\ref{sec: simulation environment} we provide a description of the simulation environment and Section~\ref{sec: simulation results} reports on extensive simulation experiments, in which the performance of all the policies is compared. Section~\ref{sec: conclusion} contains conclusions and some suggestions for further research.

\section{Considered Methods}
\label{sec: methods}
In this section we discuss the various AC policies that are proposed in this paper. 
Before introducing the threshold-based and reinforcement learning policies, we first discuss the reward framework. 
The purpose of this reward framework is threefold; i) it is needed for updating the Q values in the reinforcement learning policy, ii) it enables us to compare the various policies based on the total (discounted) reward, where the goal for each policy is to maximize the highest total (discounted) reward, and iii) it allows us to distinguish and prioritize between UE types. Indeed, the operator can construct the rewards in such a way that dropping a certain type is highly unfavorable relative to the other types. Let $r_{\mathrm{x},C(i)}$ be the reward of the event $\mathrm{x}$ for the $i$th UE, here $C: I \rightarrow M$ is a function that maps the UE to its type with $I$ denoting the set of UEs and $M$ the set of types. Accepting a new UE connection request yields a (positive) reward $r_{\mathrm{A},C(i)}$ while for blocking this UE connection we pay (negative) penalty $r_{\mathrm{B},C(i)}$. Moreover, for dropping an existing UE connection we pay a penalty, i.e., receive a negative reward $r_{\mathrm{D},C(i)}$. From an user experience perspective it is more bothersome when the connection is abruptly terminated than not being able to establish the connection in the first place. 

\subsection{Threshold policies}
\label{sec: threshold policies}
We consider two threshold-based policies which are the prior art, see for example~\cite{CBVCA-NSGRS}. In the \textit{threshold UE} policy a new UE connection request is accepted if the total number of UEs served by the base station is less than a certain specified threshold $\tau_{\mathrm{U}}$. 
In the \textit{threshold resource} policy a new UE connection request is accepted if the total occupied resource of the base station plus the requested resource of the new UE connection is less than a certain specified threshold $\tau_{\mathrm{R}}$. Equivalently, this policy ensures that the fraction of occupied resources after accepting the UE connection does not exceed $\tau_{\mathrm{R}}/B$, with $B$ denoting the total bandwidth of the base station. Observe that in both policies the number of UEs or the total occupied resource can exceed the threshold in a specific base station at some point in time, since we allow for mobility of the UEs.

\subsection{Q-learning}
Q-learning is a reinforcement learning algorithm that learns the quality of actions under a generic set of circumstances, captured in the system state \cite{WD-QL}. Note that this algorithm is model-free, meaning that it does not require the formalization of any underlying model.
 
Let the action space be denoted by $\mathcal{A}$ and the state space by $\mathcal{S}$. At each execution time the AC agent can perform two actions: either block or accept new UE connection requests, i.e., $\mathcal{A} = \{\text{block}, \text{accept}\}$ which we denote by $a^{-}$ and $a^{+}$, respectively. The set of features defining the state space will be varied to investigate the effect of additional features on the performance of the agent. 
Let $s_{i} \in \mathcal{S}$ denote the state upon request of a new connection by UE $i$. For this UE connection the AC agent will take the action $a_{i} \in \mathcal{A}$ which has the highest Q value, i.e., $a_{i} = \text{arg}\max_{a \in \mathcal{A}} Q(s_{i},a)$. 

The Q values for each action are stored in a so-called look-up table and are updated based on the reward framework.
The Q values represent the expected discounted reward and are updated according to the Bellman equation given by,
\begin{align}
\label{eq: updating Q table}
Q^{\mathrm{new}}(s_{i},a_{i}) &= Q(s_{i},a_{i}) + \alpha \Big( r(s_{i},a_{i}) \nonumber \\
& \quad + \gamma^{\Delta t} \max_{a \in \mathcal{A}}Q(s_{i+1},a) - Q(s_{i},a_{i}) \Big),
\end{align}
where $\alpha$ is the learning rate, $\gamma$ the discount factor, $\Delta t$ the time until the next decision making point and the discounted reward
\begin{align*}
r(s_{i},a_{i}) =
\begin{cases} 
r_{\mathrm{B},C(i)}  & \text{ if } a_{i} = a^{-}, \\
r_{\mathrm{A},C(i)} + \gamma^{\Delta t_{\mathrm{D},i}} r_{\mathrm{D},C(i)} \mathbbm{1}_{\text{UE $i$ dropped}}  & \text{ if } a_{i} = a^{+},
\end{cases}
\end{align*}
where $\mathbbm{1}$ denotes the indicator function and $\Delta t_{\mathrm{D},i}$ the time between accepting and dropping the connection of UE $i$.
This updating rule in Q-learning requires discrete state space $\mathcal{S}$ and action space $\mathcal{A}$. Since our state space is continuous, we quantize it. Observe that there is a trade-off between minimizing the quantization error and the inefficiency of learning, which is due to the curse of dimensionality.

\begin{figure}[H]
  \centering
  \includegraphics[width=\linewidth]{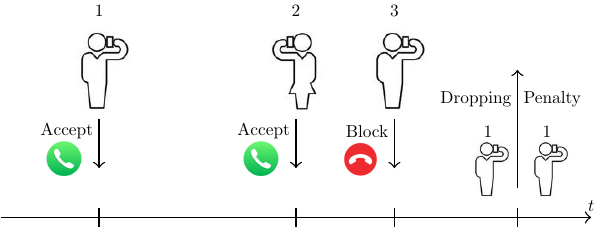}
  \caption{Visualization of the dropping rule that allocates the penalty for dropping to the state when the dropped UE was accepted.}
  \label{fig: visualization dropping}
\end{figure}

In this paper, we allocate the (negative) discounted reward of dropping $\gamma^{\Delta t_{\mathrm{D},i}} r_{\mathrm{D},j}$ to the time and state where the action of accepting that same UE was taken, see Figure~\ref{fig: visualization dropping} for the visualization of this dropping allocation.
However, one might argue that the UE that was last accepted caused the overload of the base station and therefore should be penalized or that all the UEs present at the moment of dropping are to blame. Note that in this latter case also UEs that are not dropped are penalized. We leave the comparison of all these different dropping allocations for further research. 

\subsection{Deep Q-learning}

As already mentioned above, the drawbacks of Q-learning are the discretization of the state space and the curse of dimensionality of the look-up table when increasing the number of features. 
One of the solutions to this problem is Deep Q-learning (DQL) \cite{GBC-DL}, since this approach allows for a continuous state and action space. 

\begin{figure}[h]
  \centering
  \includegraphics[width=\linewidth]{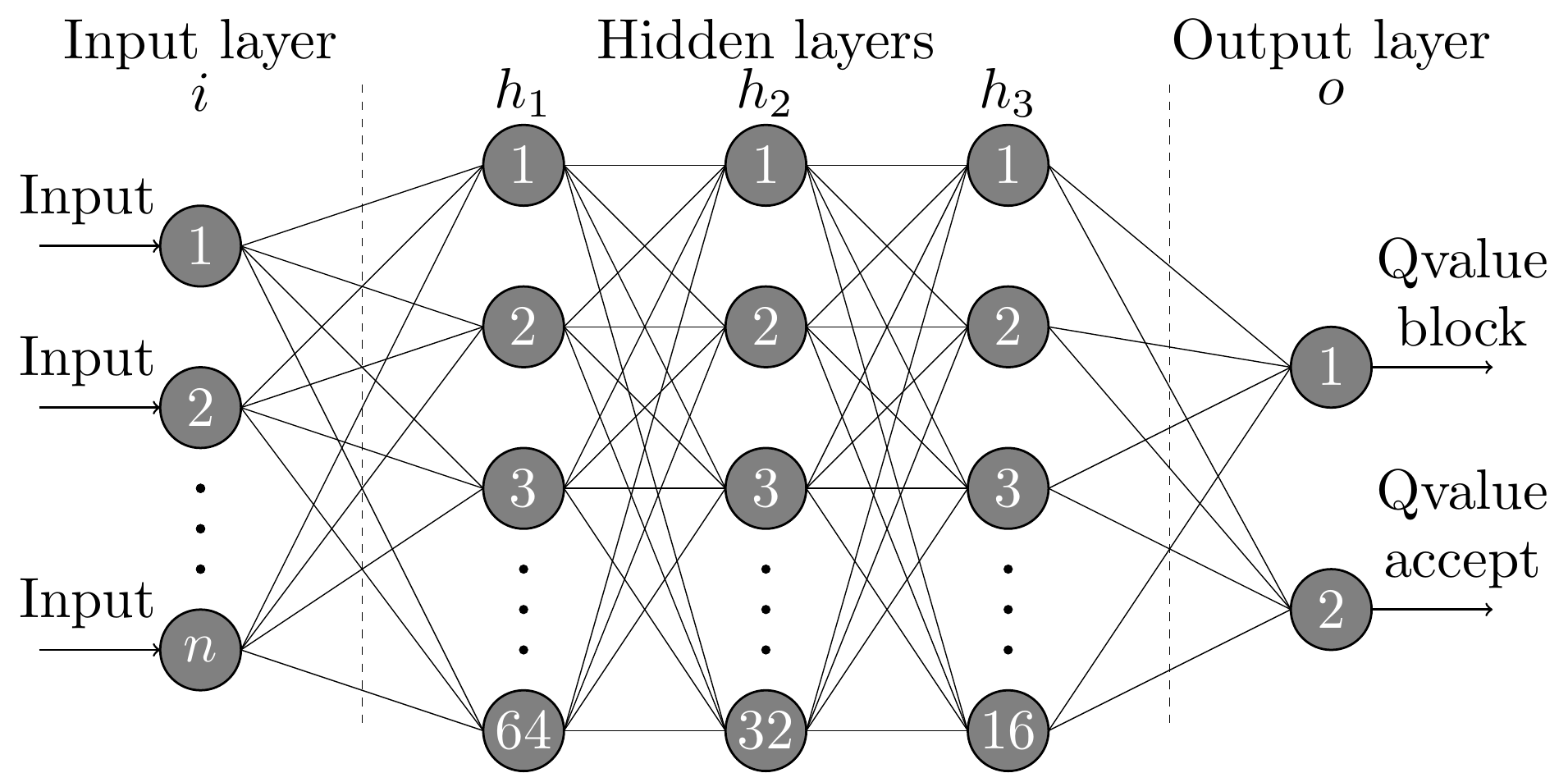}
  \caption{Visualization of the architecture of the Neural Network.}
  \label{fig: visualization neural network}
\end{figure}

In Deep Q-learning the look-up table is replaced by a (Deep) Neural Network (NN), see Figure~\ref{fig: visualization neural network} for the NN that we used for conducting the simulation experiments in Section~\ref{sec: simulation results}.

\begin{figure}[h]
  \centering
  \includegraphics[width=\linewidth]{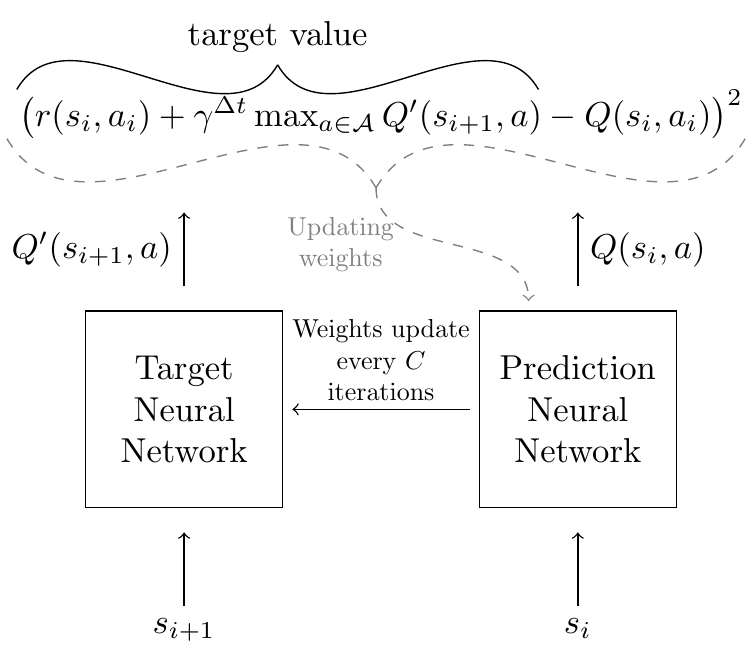}
  \caption{Visualization of the updating of the neural network.}
  \label{fig: updating neural network}
\end{figure}

Similar to the Q-learning policy, for the new UE connection request in state $s_{i}$ we take the action $a_{i}$ that gives the highest Q value, where the Q values are obtained by forwarding the input state $s_{i}$ through the Prediction Neural Network (PNN), see Figure~\ref{fig: updating neural network}. 

Updating the weights in the PNN and thus indirectly the Q values is different from the Q-learning method. Figure~\ref{fig: updating neural network} gives a schematic overview of this updating. In the figure it can be seen that besides the PNN there is also a Target Neural Network (TNN).
In Deep Q-learning we replace the value function with a function approximator. As a consequence, in Q-learning we update exactly one Q value, see Equation~\eqref{eq: updating Q table}, whereas in Deep Q-learning we update many due to the fact that all outputs depend on all neurons' weights in the network. This causes the problem that the updating affects the Q values for the next decision. Thus, the purpose of the TNN is to have a fixed neural network for the PNN to converge to, which ensures robustness and more stable convergence to the correct Q values. 

The loss of the PNN is given by,
\begin{align}
\label{eq: updating NN}
r(s_{i},a_{i}) + \gamma^{\Delta t} \max_{a \in \mathcal{A}}Q(s_{i+1},a) - Q(s_{i},a_{i}).
\end{align}
As can be seen in Figure~\ref{fig: updating neural network} we use the mean squared error (MSE) of the predicted Q-values for updating the weights of the PNN via back-propagation (with the Adam optimizer and learning rate $0.0001$). 
For the pseudo code corresponding to the DQL policy we refer to Appendix~\ref{app sec: pseudocode}.

Some comments are in place regarding the training of the DQL policy. 
The performance of this policy is highly sensitive to the hyperparameters and we observed that the DQL policy not always converged to a network that provides the right decision for accepting or blocking a new UE connection request.

\subsection{Clairvoyant}
Normally, clairvoyant policies assume to have full knowledge of the system to make optimal decisions. However, in our AC model we allow for time-varying channel rates that depend, among other parameters, on the occupied resources of each base station and in addition we allow for multiple UE types with possibly different rewards. Due to the complexity of the dependencies it is NP-hard to derive the true optimal AC policy. Next, we describe the clairvoyant policy that we consider which gives a total reward that approximates the true optimal total reward.
 
In the \textit{no dropping clairvoyant} (NDC) policy all UE connection requests are initially accepted. However, accepting all these connections may lead to overload of the base station and therefore dropping of UE connections. Whenever a UE connection is dropped, we block this UE connection request in hindsight, see Algorithm~\ref{alg: clairvoyant policy}. Note that blocking this UE connection changes the evolution of the simulation compared to the simulation in which the UE was accepted, however, we do not take this into account. This is one of the reasons why the clairvoyant policy is not truly optimal. Furthermore, observe that the dropping probability in the clairvoyant policy is zero.  

\begin{algorithm}[H]
\begin{algorithmic}[1]
\IF{new UE connection request}
\STATE Accept UE connection
\STATE Add reward for accepting the UE connection
\ENDIF
\IF{finishing UE connection}
\STATE Do nothing
\ENDIF
\IF{dropping UE connection}
\STATE Substract reward for accepting the UE connection
\STATE Add penalty for blocking the UE connection
\ENDIF
\end{algorithmic}
\caption{No dropping clairvoyant policy}
\label{alg: clairvoyant policy}
\end{algorithm}

\section{Simulation Setup}
\label{sec: simulation environment}
The considered simulation environment in this work reproduces the downlink channel between every active base station and every UE in the system, according to the 3D Urban Macro (3D-UMa) scenario defined by 3GPP~\cite{TR38.901, TR36.873}. In this work, we do not consider the effect of fast fading, but we only generate the effect of path loss, spatially coherent line/non-line of sight and shadowing, since we are interested in the average resource demand of UEs and not in stringent delay requirements. Accordingly, we assume isotropic single antenna transmitters and receivers. 

\begin{figure}[h]
\centering
\includegraphics[width=0.65\linewidth]{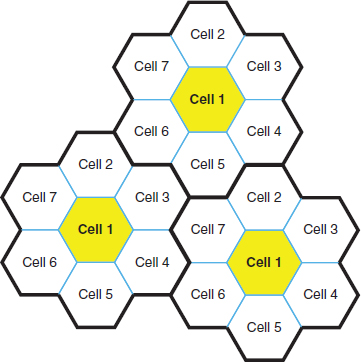}
\caption{Visualization of hexagon $7$ cell layout with wraparound.}
\label{fig: hexagon layout}
\end{figure}

We consider a hexagon $7$ cell layout with wraparound, see Figure~\ref{fig: hexagon layout}.
New UE connection requests come into the system according to a Poisson point process with rate $\lambda$ and once accepted the UE travels throughout the cells with a fixed velocity $v$ and fixed linear trajectory, determined upon acceptance. Each UE requires a certain amount of resources from the base station by which it is served. Observe that the traffic that we consider is the drop-sensitive one and that one can always operate at full capacity by also accepting best-effort connections. 

We highlight that a metric corresponds to the $i$-th UE connection with a subscript $i$. The amount of resources occupied by UE $i$ is based on 3GPP TR 38.901~\cite{TR38.901} and determined as follows. 

\begin{figure*}[htbp]
\centering
\begin{minipage}{.5\textwidth}
  \centering
    \includegraphics[width=0.9\linewidth]{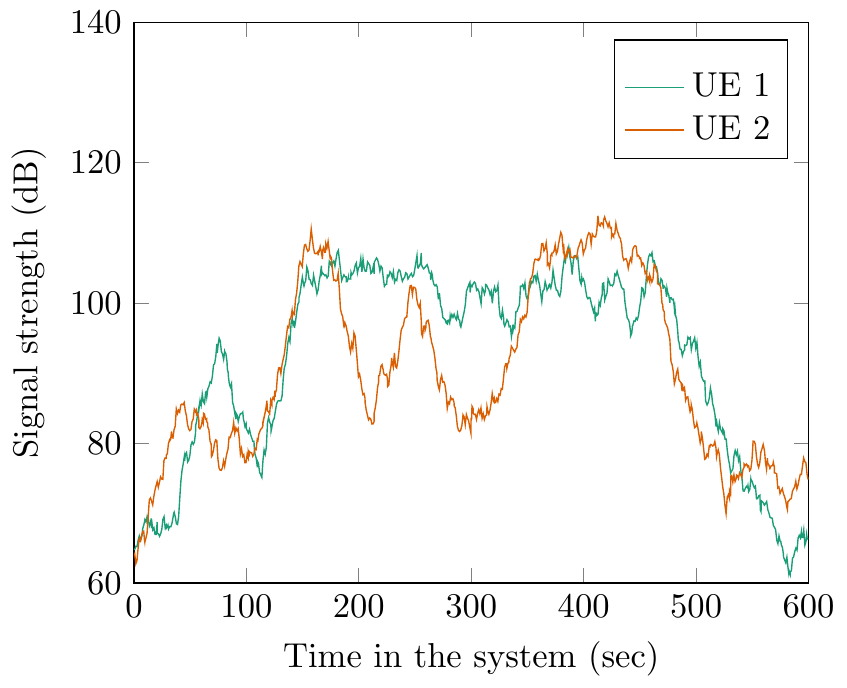}
\end{minipage}%
\begin{minipage}{.5\textwidth}
  \centering
    \includegraphics[width=0.9\linewidth]{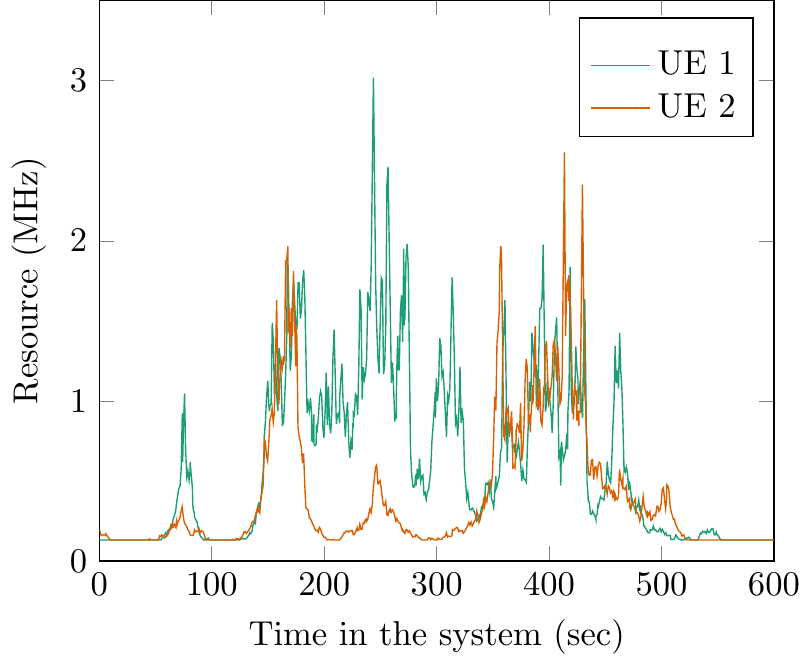}
\end{minipage}
  \caption{Realization of the signal strength and corresponding occupied resources for two UEs in the scenario with $f_{c}=2$ and $h_{\mathrm{UT},i}=1.5$. Both UEs start at the centre of cell $1$ and move with unit speed to the right according to Figure~\ref{fig: hexagon layout}.}
  \label{fig: signal strength resource}
\end{figure*}

Let $\textit{PL}_{\mathrm{LOS},i}$ denote the line of sight (LOS) pathloss of UE $i$, which is given by
\begin{align*}
\textit{PL}_{\mathrm{LOS},i} = 22 \log_{10}(d_{\mathrm{3d},i}) + 28 + 20 \log_{10} (f_{c}),
\end{align*}
where $d_{\mathrm{3d},i}$ is the 3-dimensional distance between this UE and the base station by which it is served and $f_{c}$ the carrier frequency.
Moreover, let $\textit{PL}_{\mathrm{NLOS},i}$ denote the non-line of sight (NLOS) pathloss of UE $i$ given by
\begin{align*}
\textit{PL}_{\mathrm{NLOS},i} &= 13.54 + 39.88 \log_{10}(d_{\mathrm{3d},i}) + 20 \log_{10} (f_{c}) \\
&\qquad - 0.6(h_{\mathrm{UT},i}-1.5),
\end{align*}
where $h_{\mathrm{UT},i}$ is the height of this UE. In Figure~\ref{app fig: pathloss} (Appendix~\ref{sec app: simulatin results}) both the LOS and NLOS pathloss are depicted as a function of $d_{\mathrm{3d},i}$ with $f_{c}=2$ and $h_{\mathrm{UT},i}=1.5$.

Upon acceptance of the new UE connection the UE either has LOS or NLOS pathloss with probability 
\begin{align*}
\mathbb{P}(\textit{PL}_{i} = \textit{PL}_{\mathrm{LOS},i}) = 
\begin{cases}
1 \qquad \qquad \qquad \text{ for } d_{\mathrm{2D-out},i} \leq 18,\\
\frac{18}{d_{\mathrm{2D-out},i}} + (1-\frac{18}{d_{\mathrm{2D-out},i}}) e^{- \frac{d_{\mathrm{2D-out},i}}{63}} \\ \qquad \qquad \qquad \text{ for } d_{\mathrm{2D-out},i} > 18,
\end{cases}
\end{align*}
and $1 - \mathbb{P}(\textit{PL}_{i} = \textit{PL}_{\mathrm{LOS},i})$, respectively,  
where $d_{\mathrm{2D-out},i}$ is the distance in the horizontal plane between the base station by which the UE is served and the outer wall of the building in which the UE is located, see also~\cite[Figure~7.4.1-2]{TR38.901}.
We assume that after $d_{\mathrm{cor}}$ meters the pathloss of the UE is completely uncorrelated, meaning that again the UE has $\textit{PL}_{\mathrm{LOS},i}$ or $\textit{PL}_{\mathrm{NLOS},i}$ with probability $\mathbb{P}(\textit{PL}_{i} = \textit{PL}_{\mathrm{LOS},i})$ and $1-\mathbb{P}(\textit{PL}_{i} = \textit{PL}_{\mathrm{LOS},i})$, respectively. In between, we apply linear interpolation to compute the pathloss $\textit{PL}_{i}$ of the UE $i$.  

Let $X_{i}$ denote the shadowing component of UE $i$ which has auto-correlation function
\begin{align*}
R(\Delta x) = e^{-\frac{\Delta x_{i}}{d_{\mathrm{cor}}}},
\end{align*}
where $d_{\mathrm{cor}}$ denotes the correlation length and $\Delta x_{i}$ the distance in the horizontal plane between the current position of the UE and the location upon establishing the connection. In Figure~\ref{fig: signal strength resource} (left) a realization of the signal strength is depicted, i.e., the sum of the pathloss and shadowing component ($PL_{i} + X_{i}$). 

For deriving the signal-to-interference-plus-noise ratio (SINR) of UE $i$, denoted by $\textit{SINR}_{i}$, we first have to derive the received power, the noise power and the interference power. 
The received power of UE $i$ is given by\begin{align*}
P^{R}_{i} = P^{T} \frac{B_{i}}{\textit{PL}^{j}_{i} X^{j}_{i}},
\end{align*}
where $B_{i}$ is the allocation fraction of resources to UE $i$.
The noise power is given by
\begin{align*}
P^{N}_{i} = N_{0} B B_{i},
\end{align*}
where $N_{0} = -174$ dBm/Hz and $B$ the total bandwidth in hertz. 
The interference power is given by
\begin{align*}
P^{I}_{i} = \sum_{j \in J_{-i}} P^{T} \frac{B_{i}}{PL^{j}_{i} X^{j}_{i}} B^{j},
\end{align*}
where $PL^{j}_{i}$ and $X^{j}_{i}$ denote the pathloss and the shadowing component, respectively, of UE $i$ at base station $j$, $J_{-i}$ the set of all base stations except the one of UE $i$ and $B^{j}$ the occupied resource of base station $j$. 

Now the signal-to-interference-plus-noise ratio is given by
\begin{align*}
\textit{SINR}_{i} = \frac{P^{R}_{i}}{P^{N}_{i} + P^{I}_{i}}
\end{align*}

The channel rate of the UE is given by
\begin{align*}
c_{u,i} = \log_{2}(1+\textit{SINR}_{i}),
\end{align*}
which is in the simulation capped between $0.32$ and $7.6$ which corresponds to $-6$ and $22.9$ in dB, respectively. 
The resources needed of the base station $j_{i}$ for UE $i$ is equal to $\gamma_{\mathrm{T},i} / c_{u,i}$, where $\gamma_{\mathrm{T},i}$ is the  minimum rate requirement of UE $i$. The base station that serves the UE is changed when another base station has better combined signal strength with a margin of $3$ dB. 
For a realization of the occupied resource of an UE see Figure~\ref{fig: signal strength resource} (right) (the corresponding channel rate can be found in Appendix~\ref{sec app: simulatin results}, Figure~\ref{app fig: channel rate}).

After a UE connection has been accepted, it either finishes service or it is dropped because the total occupied resources at the base station by which it is currently served exceed the total bandwidth. In this paper we consider the \textit{cost per resource} dropping rule. As the name suggests, the UE connection at the base station of interest that has the lowest cost, i.e. the smallest penalty, per resource is terminated. Other dropping rules might include:
\begin{enumerate}[i)]
\item \textit{random} in which a random UE connection is terminated,
\item \textit{channel rate} in which the UE connection with the lowest channel rate is terminated,
\item \textit{last acceptance} in which the UE that established the connection the latest is terminated.
\end{enumerate}
We leave the comparison of different dropping rules for further research.

An overview of all the input parameters can be found in Table~\ref{tab: input parameters}. 
\begin{table}[h]
\caption{Input parameters for the simulation.}
\label{tab: input parameters}
\centering
\begin{tabular}{|l|l|}
\hline
Arrivals as Poisson point process & $\lambda$ \\ \hline
Inter Site Distance & $400$ m \\ \hline
Velocity UEs & $v \in \mathrm{Unif}[1,5]$ m/s \\ \hline
Holding time of UEs & $X_{\mathrm{B}} \in \mathrm{Exp}(\mu=0.005)$ s \\ \hline
Carrier frequency  & $f_{c}=2$ GHz \\ \hline
Transmit power & $P^{T}=46${ dBm} \\ \hline
Base station height & $h_{\mathrm{bs}}=25$ m \\ \hline
UE height & $h_{\mathrm{UT}}=1.5$ m \\ \hline
Shadowing component & $X_{\mathrm{s}} \in \mathcal{N}(0,\sigma_{\mathrm{s}}^{2})$ with $\sigma_{\mathrm{s}}=4$ dB \\ \hline
Distance correlation for shadowing & $d_{\mathrm{cor}}=37$ m \\ \hline
Throughput & $\gamma_{\mathrm{T}}=1$ Mb/s \\ \hline
Total bandwidth cell & $B = 10$ MHz \\ \hline
\end{tabular}
\end{table}

\begin{figure}[ht]
\centering
\includegraphics[width=\linewidth]{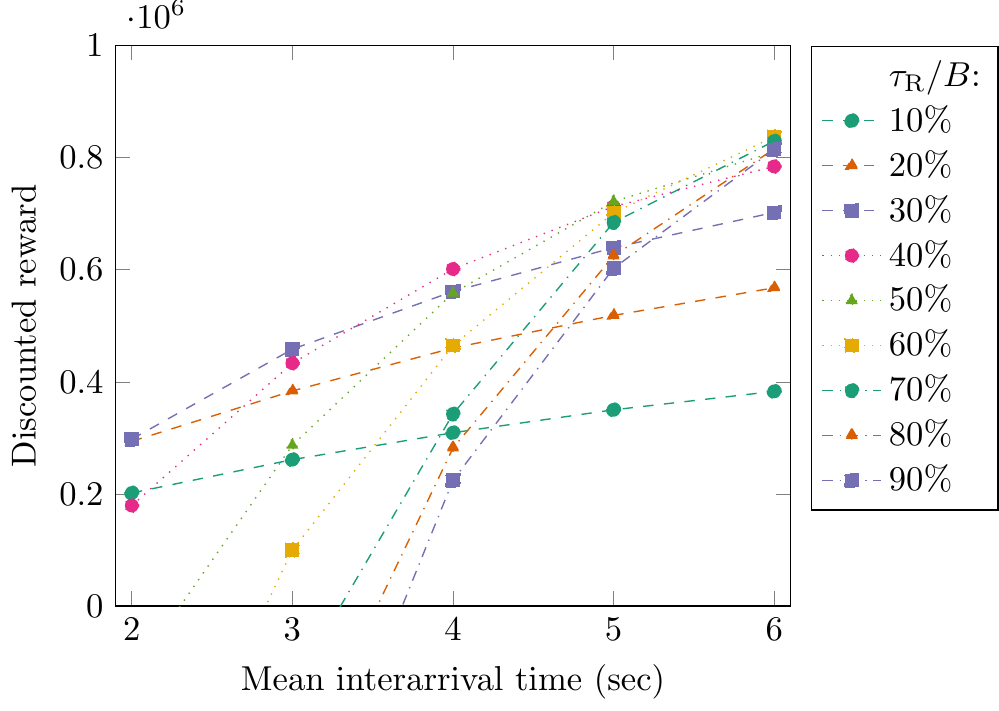}
  \caption{Discounted reward for the threshold policy based on the occupied resource of the base station for several threshold values $\tau_{\mathrm{R}}/B$ in the scenario with homogeneous arrivals, one UE type with rewards $r_{\mathrm{A}} = 10$, $r_{\mathrm{B}} = 0$ and $r_{\mathrm{D}} = -100$, and cost per resource dropping.}
  \label{fig: discounted cost threshold FullCell types1 seed1 CPR}
\end{figure}

\begin{figure*}[htbp]
\centering
\includegraphics[width=\textwidth]{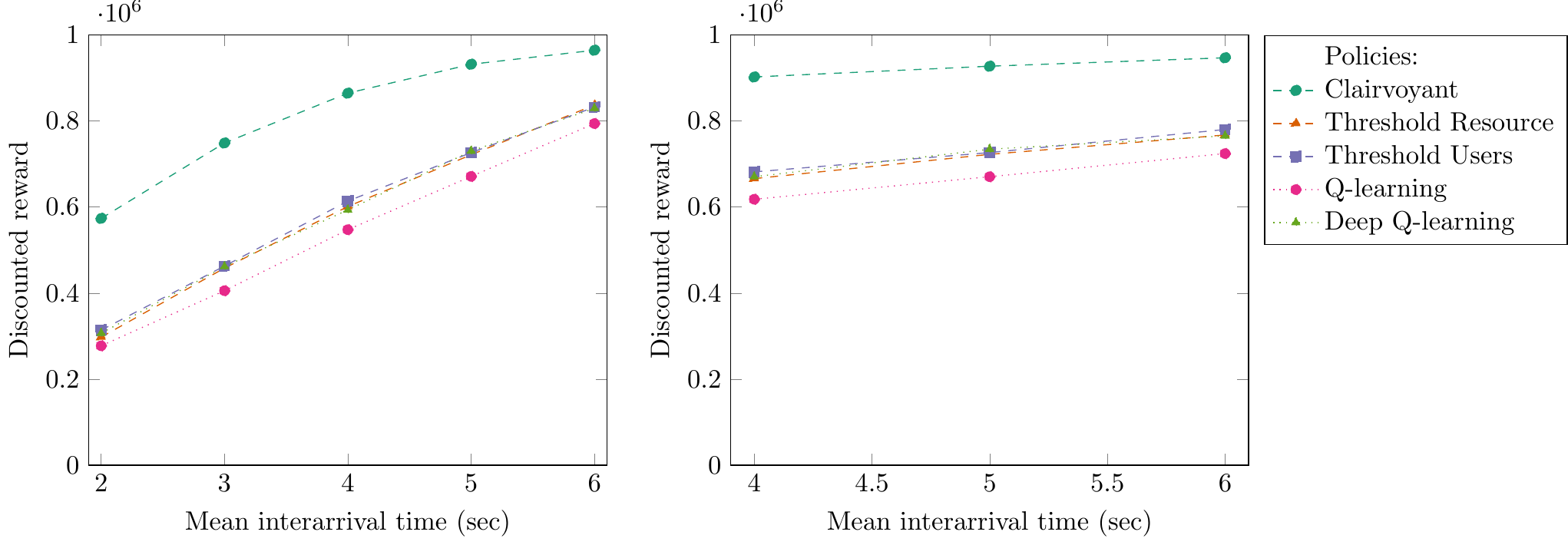}
  \caption{Discounted reward for various policies for the scenario with homogeneous (left) and heterogeneous (right) arrival rates, one UE type with rewards $r_{\mathrm{A}} = 10$, $r_{\mathrm{B}} = 0$ and $r_{\mathrm{D}} = -100$, and cost per resource dropping. Note that both the QL and DQL policy are trained to include the arrival rate (as input).}
  \label{fig: discounted cost ALL FullCell types1 seed1 fixed rates}
\end{figure*}
\section{Simulation results} 
\label{sec: simulation results}
In this section we study the performance of the AC policies described in Section~\ref{sec: methods} in various scenarios. First, in Sections~\ref{sec: results threshold} and \ref{sec: results NN}, we will consider a scenario in which the rates for new UE connection requests are spatially homogeneous, meaning that the rates in all cells are equal. Moreover, we conduct every simulation with $10^{5}$ UE connection requests and consider the \textit{cost per resource} dropping rule where the (negative) penalty of dropping is allocated to the UE that is dropped.

\subsection{Threshold policy}
\label{sec: results threshold}

In Section~\ref{sec: methods} we distinguish between two threshold policies. In Figure~\ref{fig: discounted cost threshold FullCell types1 seed1 CPR} the threshold policy based on the occupied resources is depicted for various threshold values (the threshold policy based on the number of UEs in the cell shows similar behavior). It can be seen that the best threshold value differs depending on the rate of the UE connection requests. 

Interestingly, Figure~\ref{fig: discounted cost threshold FullCell types1 seed1 CPR} shows that the best resource based threshold policy accepts a UE connection when the load of drop-sensitive traffic is between 40\% and 60\% depending on the rate. Note that, due to the time-varying channel rates, the fraction of occupied resources of this traffic may exceed the specified threshold value as already mentioned at the introduction of the threshold based policies. 

Observe that the discounted reward is increasing with the mean interarrival time, since a larger mean interarrival time means that within a certain time interval there are fewer UE connection requests. Fewer UE connection requests per time interval in turn means that we can accept a higher fraction of the UE connection requests as there will be fewer UE connections served by the base station. 

In the remainder of this paper we will only depict the \textit{frontier} of both threshold policies, i.e., the highest discounted reward for each mean interarrival time. 

\subsection{Deep Q-learning policy}
\label{sec: results NN}
For the Deep Q-learning policy we will first study the performance of having additional features describing the state space.
The versions, which represent a different set of features that are used as state space description, can be summarized as follows, where in each version the following feature(s) are added:
\begin{enumerate}[1)]
\item total resource usage of the cell,
\item resource usage of the arriving UE,
\item total resource usage of the neighboring cells,
\item quality of the UEs in the cell.
\end{enumerate}

These features are carefully chosen to potentially increase the performance, but also to represent a practical implementation of the policy. For example, adding features such as the trajectory and velocity of the UE might increase the performance but this is achieved by over-fitting the simulation model. The aim of our AC policy is to employ it in wireless networks in which the trajectory and velocity of an UE are in most cases unknown. Note that the UE type is included as a feature using one-hot encoding.    


Surprisingly, simulation experiments (not included) demonstrated that all the versions perform equally well.
One would think that adding, for example, the total resource usage of neighboring cells as a feature would improve the DQL policy. However, observe that the state space does not include the trajectory and speed of all the UEs. Consider the situation where one neighboring cell is fully loaded while all other neighboring cells are empty. On the one hand, if the DQL policy accepts a new UE connection there is the probability of $1/7$ that the UE goes to the cell that is fully loaded leading to a dropping. On the other hand, if the DQL policy blocks the new UE connection it is not using the full capacity of the base station corresponding to this UE. In addition, simulation experiments showed that the fraction of occupied resources at the base station is highly fluctuating.

\subsection{Comparison of various policies}
We compare the performance of the various policies for three different scenarios; spatially homogeneous (equal rates in all cells), heterogeneous (different rates in the cells) and time-varying heterogeneous arrival rates. The time-varying arrival rates are modeled by having every $t_{\text{var}}$ seconds new heterogeneous arrival rates in the cells. Note that for both threshold policies, i.e., based on the number of UEs and on the occupied resource, only the frontier is depicted. For the QL and DQL policies we trained the agent in such a way that it is able to cope with general arrival rates, meaning that we have a look-up table and NN for the QL and DQL policy, respectively, in which the state space includes the arrival rate.

Figure~\ref{fig: discounted cost ALL FullCell types1 seed1 fixed rates} shows that the DQL policy is performing equally good as the frontier of the threshold-based policies both for fixed homogeneous and heterogeneous arrival rates, whereas the QL policy performs slightly worse. The lower performance of the QL policy might be due to the quantization error. 

\begin{figure}[ht]
\centering
\includegraphics[width=\linewidth]{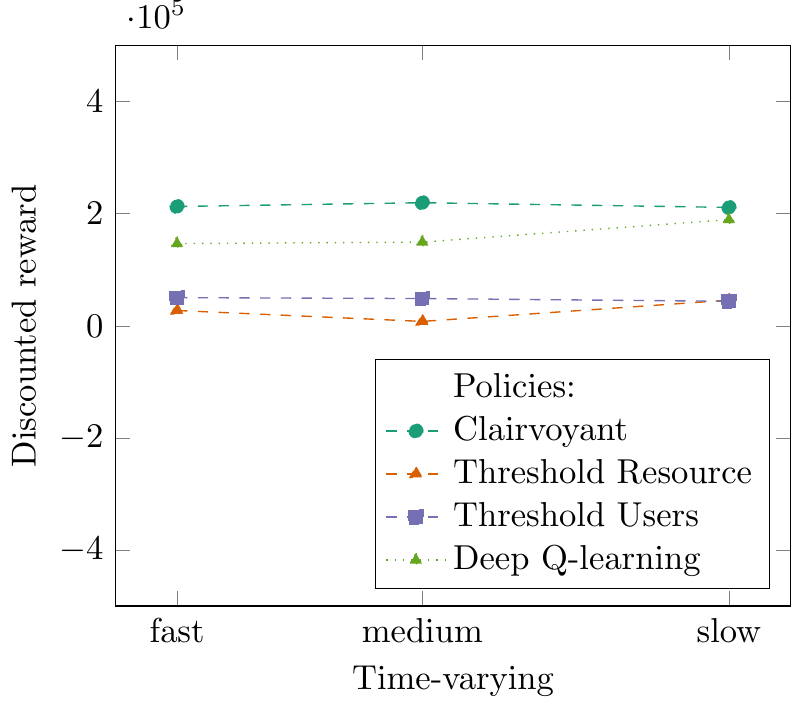}
\caption{Discounted reward for various policies for the scenario with fast ($t_{\text{var}}=1000$), medium ($t_{\text{var}}=5000$) and slow ($t_{\text{var}}=10000$) time-varying heterogeneous arrival rates, two UE types with rewards $\boldsymbol{r}_{\mathrm{A}} = (10,1)$, $\boldsymbol{r}_{\mathrm{B}} = (-10,-1)$ and $\boldsymbol{r}_{\mathrm{D}} = (-100,-10)$, and cost per resource dropping. Note that the DQL policy is trained to capture various arrival rates.}
\label{fig: discounted cost neuralnetwork FullCell types2 seed1 CPR timevarying}
\end{figure}

We cannot say with certainty if these results suggest that the DQL policy is not able to discover and/or capture system characteristics that improve the performance or that it is just impossible and not a weakness of this policy. 
Note that for the threshold policies we manually set the correct threshold value to achieve the frontier, depicted in Figure~\ref{fig: discounted cost ALL FullCell types1 seed1 fixed rates}, for a specific mean interarrival time which a practical system would first need to predict from past arrivals with according errors. Both the QL and DQL policy are more general than the threshold-based policies in the sense that these policies are able to adaptively learn the decision that achieves this performance. In addition both the QL and DQL policies are able to capture general arrival rates, whereas the best threshold value depends on the mean interarrival time and again should be set manually. 

Figure~\ref{fig: discounted cost neuralnetwork FullCell types2 seed1 CPR timevarying} shows the performance of the AC policies in the most realistic scenario, i.e., multiple UE types with heterogeneous time-varying arrival rates. For this scenario the DQL policy is outperforming the threshold-based policies. Note that in the threshold policies we manually set one threshold for every UE type (independent of the rate of new UE connection requests) once and for all in the simulation. Adapting the threshold for the various UE types is feasible but would require tuning with even more complexity. The (well-trained) DQL policy is able to capture multiple UE types and the time varying rate of new UE connection requests. Note that the DQL policy can be arbitrarily extended and trained with the same procedure, independent of the number of UE types. 

The reward framework allows us to compare the performance of various policies, but it does not provide insight in the important performance metrics within AC such as the blocking and dropping probability. In Table~\ref{tab: probabilities timevarying} these latter performance metrics are shown. Note that the scenario that we considered has one UE type with a high reward for acceptance, whereas the other UE type has a low reward for acceptance. In the table it can be seen that the threshold-based policy does not differentiate between the two UE types. In contrast, the DQL policy significantly favors the high reward UE type over the UE type with low reward. Moreover, observe that the DQL policy achieves a lower dropping probability for both UE types. According to~\cite{ITU-E.807} the desired drop rate should be at most $3\%$. Note that the dropping probability achieved by the algorithms can be increased by decreasing the penalty of dropping, and vice-versa. 

\begin{table}[h]
\caption{Acceptance and dropping probabilities for various policies in the same scenario as Figure~\ref{fig: discounted cost neuralnetwork FullCell types2 seed1 CPR timevarying} with fast ($t_{\text{var}}=1000$) varying heterogeneous arrival rates.}
\label{tab: probabilities timevarying}
\centering
\begin{tabular}{l|l|l|l|l}
 & \multicolumn{2}{l|}{UE type $1$} & \multicolumn{2}{l}{UE type $2$} \\ \cline{2-5}
 & Accept & Dropping & Accept & Dropping \\ \hline
Clairvoyant & $0.906$ & $0.0$ & $0.644$ & $0.0$ \\ \hline
Threshold resource & $0.646$ & $0.008$ & $0.646$ & $0.042$ \\ \hline
Threshold UEs & $0.724$ & $0.009$ & $0.718$ & $0.05$ \\ \hline
Deep Q-learning & $0.738$ & $0.005$ & $0.426$ & $0.023$
\end{tabular}
\end{table}

\section{Conclusion}
\label{sec: conclusion}

In this paper we studied the performance of various policies for admission control in wireless networks. For fixed load and one UE type, the results indicate that the prior art threshold policies perform equally good as our proposed Q-learning (QL) and Deep Q-learning (DQL) policies. However, note that both the QL and DQL policy adaptively learn the decision, whereas in the threshold-based policies we have to manually set the correct threshold which is highly impractical. 
For time-varying heterogeneous arrival rates and multiple UE types the DQL policy outperforms the threshold-based policies, due to the ability of capturing the generality of the arrival rate and multiple UE types.

For further research one could investigate additional features for the DQL policy that were not considered in this paper.  
Another extension in the DQL policy would be to add a long-short term memory (LSTM) cell in the NN, see~\cite{HS-LSTM}. 
Lastly, it is widely agreed that network slicing is a key requirement for 5G mobile networks, see for example~\cite{ABKKM-EAMSWA, CBVCA-NSGRS, MABK-SNSC}. Note that our models can be easily extended and adapted to cope with the network slicing framework by splitting users of different slices into different types, each with different reward policies.

\section*{Acknowledgments}
The authors gratefully acknowledge insightful discussions with Sem Borst and Thorsten Wild.

\bibliographystyle{plain}
\bibliography{references}

\appendices

\newpage

\section{Additional Simulation Comments}
\label{sec app: simulatin results}
In this appendix we provide additional simulation results and the pseudo code of the simulation as well as for updating the Deep Q-learning policy. 

\begin{figure}[H]
  \centering
  \includegraphics[width=0.9\linewidth]{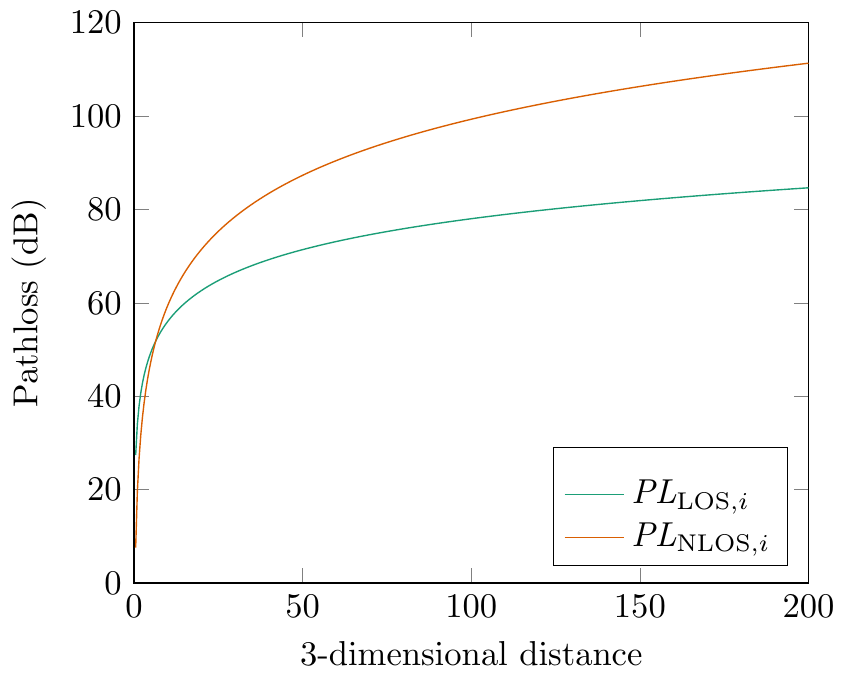}
  \caption{LOS and NLOS pathloss for UE $i$ in the scenario with $f_{c}=2$ and $h_{\mathrm{UT},i}=1.5$.}
  \label{app fig: pathloss}
\end{figure}

In Figure~\ref{app fig: pathloss} the pathloss corresponding to the expressions in Section~\ref{sec: simulation environment} are shown as function of the 3-dimensional distance $d_{\mathrm{3d},i}$. 
It can be seen that at the cell edge the difference between LOS and NLOS pathloss is approximately $30$ dB.

\begin{figure}[H]
  \centering
  \includegraphics[width=0.9\linewidth]{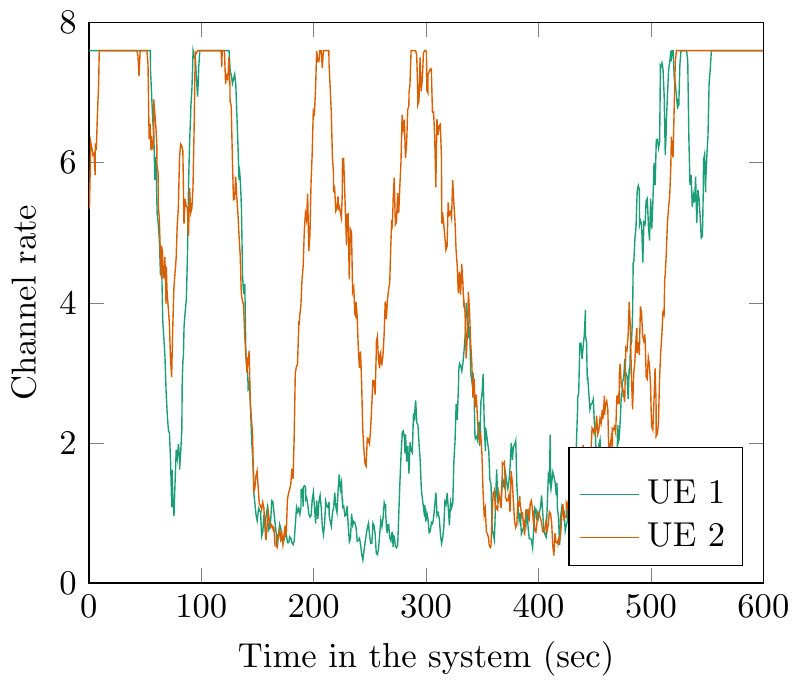}
  \caption{Channel rate of two UEs corresponding to the signal strength realization depicted in Figure~\ref{fig: signal strength resource} in the scenario with $f_{c}=2$ and $h_{\mathrm{UT},i}=1.5$. Both UEs start at the centre of cell $1$ and move with unit speed to the right according to Figure~\ref{fig: hexagon layout}.}
  \label{app fig: channel rate}
\end{figure}

Figure~\ref{app fig: channel rate} shows the channel rate of two UEs corresponding to the signal strength realization depicted in Figure~\ref{fig: signal strength resource}. It can be seen that the channel rate is highly fluctuating. Especially UE $2$ experiences steep drops in the channel rate.  

\newpage

\section{Pseudo code}
\label{app sec: pseudocode}
\subsection{Simulation}
\label{app sec: simulation}
\begin{algorithm}[H]
\floatname{algorithm}{Simulation}
\renewcommand{\thealgorithm}{}
\begin{algorithmic}[1]
\IF{new UE connection request}
\STATE{\textbf{Update} occupied resources usage UEs}
\STATE{\textbf{Check} for dropping UE connections}
\STATE{\textbf{Accept/Block} UE connection by AC policy}
	\IF{Accept}
	\STATE{Add reward for accepting the UE connection}
	\STATE{\textbf{Update} occupied resources usage UEs}
	\STATE{\textbf{Check} for dropping UE connections}
	\ELSE
	\STATE{Add penalty for blocking the UE connection}
	\ENDIF
	\STATE{Schedule new UE connection request}
\ENDIF
\IF{finishing UE connection}
\STATE{\textbf{Update} occupied resources usage UEs}
\ENDIF
\IF{dropping UE connection}
\STATE{\textbf{Remove} UE connection}
\STATE{\textbf{Update} occupied resources usage UEs}
\ENDIF
\\
\STATE{\textbf{Every second:}}
\STATE{\textbf{Update} occupied resources usage UEs}
\STATE{\qquad \textbf{Check} for dropping UE connections}
\end{algorithmic}
\caption{Admission Control}
\label{alg: AC}
\end{algorithm}

\subsection{Updating Deep Q-learning policy}
\label{app sec: updating DQL policy}
\setcounter{algorithm}{1}
\begin{algorithm}[H]
\begin{algorithmic}[1]
\STATE{\textbf{Decision} (at new connection request $i$th UE):}
\SCOPE
\STATE{\textbf{Get} and \textbf{store} input state $s_{i}$}
\STATE{\textbf{Obtain} $Q(a,s_{i})$ by \textbf{forwarding} $s_{i}$ through the PNN}
\IF{Greedy (with probability $\boldsymbol{1-\epsilon}$)}
\STATE{\textbf{Take} action corresponding to highest Q value}
\ELSE
\STATE{\textbf{Take} random action}
\ENDIF
\ENDSCOPE
\STATE{\textbf{Updating} (finishing or dropping $i$th UE connection):}
\SCOPE
\STATE{\textbf{Get} input state $s_{i}$ of the UE}
\STATE{\textbf{Obtain} $Q(a,s_{i})$ by \textbf{forwarding} $s$ through the PNN}
\STATE{\textbf{Get} input state $s_{i+1}$ of the next UE connection}
\STATE{\textbf{Obtain} $Q'(a,s_{i+1})$ by \textbf{forwarding} $s_{i+1}$ through the TNN}
\STATE{\textbf{Calculate} the loss according to Equation~\eqref{eq: updating NN}}
\STATE{\textbf{Update} weights PNN by back-propagating loss}
\ENDSCOPE
\STATE{\textbf{Every $C$ iterations:}}
\SCOPE
\STATE{\textbf{Copy} weights from PNN to TNN}
\ENDSCOPE
\end{algorithmic}
\caption{Deep Q-learning policy}
\label{alg: NN}
\end{algorithm}

\end{document}